\documentstyle[twocolumn,epsf]{jpsj}
\newcommand{\ds}{\displaystyle}

\newcommand{\non}{\nonumber}

\newcommand{\ii}{{\rm i}}

\newcommand{\para}{\parallel}
\newcommand{\beq}{\begin{eqnarray}}
\newcommand{\eeq}{\end{eqnarray}}

\newcommand{\be}{\begin{eqnarray}}
\newcommand{\en}{\end{eqnarray}}
\newcommand{\JPSJ}{J.Phys.Soc.Jpn.}

\newcommand{\PRL}{ Phys. Rev. Lett.}
\newcommand{\PRB}{ Phys. Rev.{\bf  B}.}

\def\runtitle{Spin Gap and Superconductivity in   Weakly Coupled  Ladders  }
\def\runauthor{Jun-ichiro Kishine and Kenji Yonemitsu}
\title{Spin Gap and Superconductivity in   Weakly Coupled  Ladders: 
Interladder One-particle  vs. Two-particle Crossover}
\author{Jun-ichiro Kishine\thanks{E-mail:kishine@ims.ac.jp} and Kenji Yonemitsu}
\inst
{Department of Theoretical Studies,
Institute for Molecular Science,
Okazaki 444, Japan}
\recdate{September 17, 1997}
\abst
{Effects 
of the interladder one-particle hopping, $t_{\perp}$, on the low-energy 
asymptotics of a weakly coupled Hubbard ladder system have been studied, 
based on the perturbative renormalization-group approach.
We found that  for  finite intraladder  Hubbard repulsion, $U$, there exists a crossover value of the interladder one-particle hopping, $t_{\perp c}$.
For $0<t_{\perp}<t_{\perp c}$,  the spin gap metal (SGM) phase of the isolated ladder  transits at a finite transition temperature, $T_{c}$, 
to the $d$-wave superconducting (SCd) phase  via a two-particle crossover.
In the temperature region, $T<T_{c}$, interladder {\it coherent   Josephson tunneling} of 
the Cooper pairs occurs, while
 the  interladder coherent  one-particle process is strongly suppressed.
For $t_{\perp c}<t_{\perp}$, around a crossover temperature, 
$T_{\rm cross}$, the system crosses over to the   two-dimensional (2D)
phase  via a one-particle crossover. In the temperature region,  $T<T_{\rm cross}$, the interladder{\it coherent}
 band motion occurs.
}
\kword
{doped Hubbard ladders, interladder coupling, one-particle  crossover, 
two-particle crossover, $d$-wave superconductivity,
perturbative renormalization-group }

\begin{document}
\maketitle
\baselineskip12pt
Last year Uehara {\it et al}.\cite{AkimitsuG}   discovered the 
superconductivity signal with $T_{c}=12$K in the doped spin ladder, 
Sr$_{0.4}$Ca$_{13.6}$Cu$_{24}$O$_{41.84}$, under a  pressure of 3GPa.
The electric properties of the compound are determined by the hole-doped 
ladders instead of chains.\cite{UchidaG}
A remarkable feature of the doped ladder is the existence of a spin 
excitation gap.\cite{Magishi}
The superconducting transition  under a high pressure suggests that 
interladder one-particle hopping induced by the applied pressure
plays an important role. 
Recent experiments on the resistivity along the ladder, $\rho_{c}$, of the single crystal 
Sr$_{2.5}$Ca$_{11.5}$Cu$_{24}$O$_{41}$\cite{mottAkimitsu} 
shows that  the  superconductivity  sets in below 10 K under 3.5GPa $\sim$ 8GPa with the 
temperature dependence of $\rho_{c}$ changing gradually from
$T$-linear to $T^2$.
The anisotropy of the resistivity
also indicates the dimensional crossover from 1D to 2D with increasing an applied pressure.

In this paper, to elucidate  the nature of  superconductivity in the doped  
ladder compound under  pressure, 
 we consider Hubbard ladders\cite{comment} coupled via a weak interladder
one-particle hopping.
In the case of  the isolated Hubbard ladder, the most relevant phase 
 is characterized by a strong coupling fixed point and is denoted by  
{\lq\lq}phase I{\rq\rq} by 
Fabrizio\cite{Fabrizio} and {\lq\lq}C1S0 phase{\rq\rq} by Balents and 
Fisher.\cite{BF}
In this phase,   only the total charge mode remains gapless\cite{Fabrizio,BF,KR,Schulz} and consequently, 
the $d$-wave superconducting correlation  becomes the most dominant one as long as the 
intraladder correlation is weak.\cite{Fabrizio,BF,KR,Schulz}
From now on we call this strong coupling phase {\lq\lq}spin gap metal (SGM) phase{\rq\rq}.
So far, the effects of interladder hopping on the SGM phase have 
been studied through   mean field approximations\cite{tsunetsugu,OrignacGiamarchi}
and   power counting arguments.\cite{BF2}

The central problem here is how
a  weak interladder 
one-particle hopping, $t_{\perp}$, affects the low-energy 
asymptotics of the system.
 Based on the perturbative renormali\-zation-group (PRG)
approach, we here study  one-particle and two-particle crossovers,\cite{Bourbonnais} 
when we switch on $t_{\perp}$ and the intraladder Hubbard  repulsion, 
$U$, as perturbations to the system, specified by the intraladder 
longitudinal (transverse) hopping, $t$($t'$): $U,t_{\perp}\ll t,t'$.
A similar approach has been considered for the problem of   a coupled chain 
system by Boies {\it et al}.\cite{Boies}
The intraladder one-particle process is
 diagonalized in terms of the bonding ($B$) and antibonding ($A$) bands.
As shown in Fig.~1(a), we linearize the dispersions 
along the legs on the 
bonding and antibonding Fermi points, $\pm k_{Fm}$($m=A,B$). In Fig.~1,  $R$ and $L$ 
denote the right-going and left-going branches, respectively.
The Fermi velocities  in principle depend on the band index as
$v_{Fm}=2t\sin k_{Fm}$
but we  assume throughout this work that $v_{Fm}=v_{F}$ and drop the band index, since
 the  difference in the  Fermi velocities does not affect the asymptotic 
 nature of the SGM phase at least for small $t'/t$.\cite{Fabrizio,BF}
In all  the four branches ($LB,LA,RA,RB$) of  linearized bands,  the 
energy variables, $\varepsilon_{\nu m}$ ($\nu=R,L$ ; $m=A,B$) run over the region,
$-E/2<\varepsilon_{\nu m}<E/2$, with
$E$ denoting the {\it bandwidth cutoff}.

The intraladder Hubbard repulsion generates the scattering processes depicted in Fig.~1(b).
The processes are specified by dimensionless coupling constants $g^{(1)}_{\mu}$ and  $g^{(2)}_{\mu}$  
denoting backward and forward scatterings, respectively,
with the flavor indices\cite{Fabrizio} $\mu=0,f,t$ denoting intraband scattering, interband forward scattering
 and interband tunneling processes, respectively. The usual coupling constants with a
 dimension of the interaction energy are $2\pi v_{F}g_{\mu}^{(i)}$.
We neglect the  interband backward  processes such as  
$(RB\to RA;LA\to LB)$
on the grounds that  
these processes  do not seriously modify  the asymptotic nature of the 
SGM phase.\cite{Fabrizio,BF}
\begin{figure}
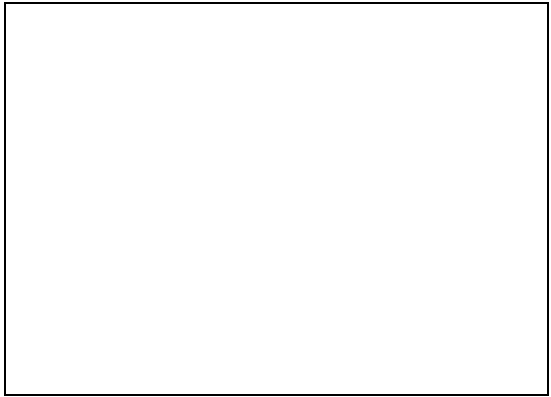

\begin{center}
\figureheight{5cm}
\caption{(a) Four branches($LB,LA,RA,RB$) of  linearized bands with the bandwidth cutoff $E$ and 
(b) intraladder two-particle scattering vertices $g^{(i)}_{\mu}$. The 
solid and broken lines represent the propagators for the right-moving and 
left-moving electrons. $m$ and $\bar m$ denote different bands.}
\end{center}
\end{figure}

When we switch on $g_{\mu}^{(i)}$ and $t_{\perp}$ as perturbations to the 
system with
$t$ and $t'$, and the temperature scale decreases, two kinds of dimensional crossovers, 
a one-particle crossover and a two-particle crossover,\cite{Bourbonnais,BY} 
occur. We illustrate one- and two-particle processes in Fig.~2.
In the  case where the interladder one-particle hopping  modified by the intraladder self-energy 
 becomes the most relevant,  a particle hops coherently from
one ladder to a neighboring one  and the system crosses over to a 
two-dimensional system via the {\it one-particle crossover}. 
In the case where the interladder two-particle process {\it generated} by 
the {\it inter}ladder one-particle 
hopping and {\it intra}ladder two-particle scattering 
processes becomes the most  relevant,  a pair of composite particles hops coherently from one ladder to a 
neighboring one and the system crosses over to a long-range-ordered phase 
via the {\it two-particle crossover}.  
In Fig.~2, we show only the   two-particle hopping in the 
$d$-wave superconducting channel, which   corresponds to the 
interladder  Josephson tunneling of   Cooper pairs.
\begin{figure}
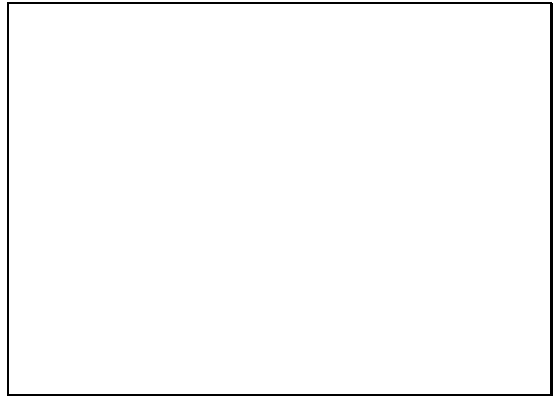

\begin{center}
\figureheight{5cm}
\caption{Schematic illustrations of the one-particle and the  two-particle process (in the case 
of $d$-wave superconductivity channel). 
In the one-particle process, a particle  hops  from one ladder to a 
neighboring one, while
in the two-particle process,
 a pair of particles hops  from one ladder to a 
neighboring one.}
\end{center}
\end{figure}

To study the competition between the one-particle crossover and the two-particle crossover,  we set up   scaling 
equations for the interladder one-particle  and two-particle 
hopping amplitudes and study their low-energy asymptotic behavior.
We start with the path integral representation of the partition function 
of the system,
$
Z=\ds\int{\cal D} e^{S},
$
where   the action  consists of four parts,
\beq
S=S^{(1)}_{\para}+S^{(2)}_{\para}+S^{(1)}_{\perp}+S^{(2)}_{\perp},\label{eqn:action}
\eeq
where 
$S^{(1)}_{\para}$, $S^{(2)}_{\para}$, $S^{(1)}_{\perp}$ and 
$S^{(2)}_{\perp}$ denote the action 
for the intraladder one-particle hopping, intraladder two-particle 
scatterings,  interladder 
one-particle and interladder two-particle hopping, respectively. 
 $\cal D$  symbolizes the measure of the path integral over the fermionic Grassmann variables.

The idea of scaling is to eliminate  the short-wavelength degrees of 
freedom  to relate   
 effective actions  at successive energy   scales.
Based on the bandwidth cutoff regularization scheme, 
we parametrize the cutoff as $E(l)=E_{0}e^{-l}$ with the scaling parameter, $l$,
and study how  the  action will be renormalized as $l$ goes from zero to infinity.
The scaling equations for all the processes studied here are depicted in Fig.~3.
\begin{figure}
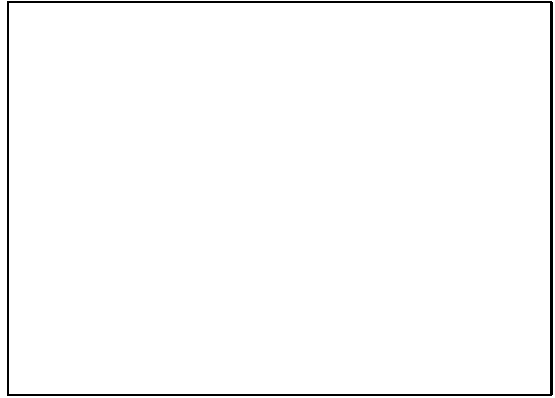

\begin{center}
\figureheight{5cm}
\caption{Diagrammatic representations of the scaling equations for the 
intraladder scattering vertices(a), intraladder one-particle 
propagator(b), interladder one-particle hopping amplitude(d), and 
interladder two-particle hopping amplitude(d). 
A black circle, zigzag line, and shaded square represent  the {\it intra}ladder two-particle scattering 
processes (a combination of $g_{\mu}^{(i)}$), the {\it inter}ladder one-particle hopping amplitude, $t_{\perp}$, 
and the {\it inter}ladder two-particle hopping amplitude, $V^{{\rm SCd}}$ respectively. 
}
\end{center}
\end{figure}
Diagrammatic expansions for the intraladder scattering processes up to the 
3rd order   are given in Fig.3~(a). 
After taking account of the field rescaling procedure which originated from   the scaling of the intraladder 
one-particle propagators,\cite{Solyom} given in Fig.~3(b), we obtain the appropriate 
scaling equations for $g_{\mu}^{(i)}$.
Full expressions of the scaling equations for $g^{(i)}_{\mu}$   are found 
 by setting  $g_{b}^{(1)}=g_{b}^{(2)}=0$ in Eq.(A5) of ref.[6].
Starting with the Hubbard-type initial condition
\beq
 g^{(i)}_{\mu}(0)=\tilde U\equiv {U/ 4\pi v_{F}}>0,
\eeq
the scaling equations give the fixed point
\be
\begin{array}{ccc}
g^{(1)\ast}_{0}=-1,&g^{(1)\ast}_{f}=0,&g^{(1)\ast}_{t}=1,\\
g^{(2)\ast}_{0}=-{3-2\tilde U \over 4},&g^{(2)\ast}_{f}={1+2\tilde U\over 4}, & g^{(2)\ast}_{t}=1.
\end{array}\label{FP}
\en
Henceforth, we take $\tilde U=0.3$ for illustration.
For any $\tilde U>0$, the results are similar to those shown below.
We show the scaling flows for $g_{\mu}^{(i)}$ in Fig.~4(a) as functions 
of the scaling parameter, $l$.
These  flows scale the isolated Hubbard ladder to the SGM 
phase.\cite{Fabrizio,BF,KR}

The one-particle crossover is determined through the scaling equation  
\begin{eqnarray}
{d\ln t_{\perp}(l)\over dl}
&=&  
1
-({{ g}_{0}^{(1)2}}
    +{{ g}_{0}^{(2)2}}
    +{{ g}_{f}^{(1)2}}
    +{{ g}_{f}^{(2)2}}
    +{{ g}_{t}^{(1)2}}
    \non\\
    &+&
   {{ g}_{t}^{(2)2}}
  -{ g}_{0}^{(1)}
     { g}_{0}^{(2)}
    -{ g}_{f}^{(1)}
     { g}_{f}^{(2)}
    -{ g}_{t}^{(1)}
     { g}_{t}^{(2)}),\label{eqn:RGfort}
\end{eqnarray}
which is represented by Fig.~3(c). 
It is seen from (\ref{eqn:RGfort}) together with the fixed point  (\ref{FP}) that 
\beq
{d\ln t_{\perp}(l)/ dl}\stackrel{l\to\infty}{\longrightarrow}-\tilde U^{2}/2-{7/8},\label{eqn:DLcase}
\eeq
and consequently 
$t_{\perp}(l)$ {\it becomes  always irrelevant at the final stage of the scaling procedure}.
However,  $t_{\perp}(l)$   grows at an early stage of scaling since the 
intraladder couplings do not grow sufficiently as yet  (see Fig.~4).
Once  $\tilde t_{\perp}(l)\equiv t_{\perp}(l)/E_{0}$  attains an order of  unity, the weakly coupled ladder picture   breaks down and
the system is scaled to a "two-dimensional"  system via the 
one-particle crossover.\cite{Bourbonnais}
In the upper planes of Figs.~4(b) and 4(c), we show the   scaling flows of 
$\tilde t_{\perp}(l)$  with the initial conditions,
$\tilde t_{\perp}(0)\equiv \tilde t_{\perp0} =0.01$ and 0.04, respectively.
\begin{figure}
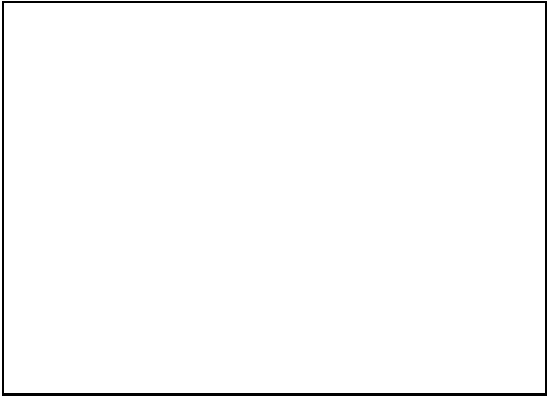

\begin{center}
\figureheight{5cm}
\caption{Scaling flows of 
$g_{\mu}^{(i)}$(a), $\tilde t_{\perp}(l)$(the upper half planes of (b),(c)), and $V^{\rm SCd}$(the lower half planes of (b),(c)).
 For the region, $l>{\rm  Min}(l_{\rm c},l_{\rm cross})$, the scaling 
 flows, denoted by  broken curves, have no physical meaning, since the weak coupling picture breaks down in the 
region.}
\end{center}
\end{figure}
We see that for $\tilde t_{\perp0}=0.01$, $\tilde t_{\perp}(l)$ never reaches unity and 
one-dimensional crossover  never takes place, while 
for $\tilde t_{\perp0}=0.04$,  $\tilde t_{\perp}(l)$  
exceeds  unity at the  scaling parameter $l_{\rm cross}$ which is defined by
\beq
\tilde t_{\perp}(l_{\rm cross})=1.\label{eqn:defoflc1}
\eeq 
Thus we   specify the one-dimensional crossover by $l_{\rm cross}$, 
although it merely has  qualitative meaning.

Interladder two-particle processes
are  decomposed into CDW, SDW, SS (singlet superconductivity) 
and TS (triplet superconductivity) 
channels as in the case of the coupled chains.\cite{Bourbonnais} In  this case, there are additional flavor indices, $\mu=0,f,t$ for each 
channel. 
Then the
two-particle hopping amplitudes  are specified as 
$V_{\mu}^{M}$ ($M$=CDW, SDW, SS, TS; $\mu=0,f,t$).
The SS channel can be decomposed into the $s$-wave spuerconductivity (SCs)   and 
$d$-wave superconductivity (SCd)
channels.\cite{Fabrizio,BF,KR,Schulz}
For the SCd channel, the action for the interladder 
two-particle hopping is written in the form
\beq
S_{\perp\rm SCd}^{(2)}=
-{\pi v_{F}\over 2\beta}\sum_{Q}{ V}^{\rm SCd}{\cal O}_{{\rm 
SCd}}^{\ast}(Q){\cal 
O}_{{\rm SCd}}(Q).
\label{eqn:V}
\eeq
The SCd  pair-field is given by
${\cal O}_{{\rm SCd}}(Q)={\cal O}_{\rm SS}^{BB}(Q)-{\cal O}_{\rm 
SS}^{AA}(Q)$, where
${\cal O}^{mm'}_{\rm SS}(Q)$  $=\beta^{-1/2}\sum_{K,\sigma}\sigma$  $R_{m,\sigma}(-K+Q)L_{m',-\sigma}(K)$,
with  $R_{m,\sigma}$($L_{m,\sigma}$) being the Grassmann variable 
representing  the right(left)-moving electron in
the  band $m$ with spin $\sigma$ and  $K=(k_{\para},k_{\perp},\ii \varepsilon_{n})$ and 
$Q=(q_{\para},q_{\perp},\ii \omega_{n})$ with
the momentum along the leg and rung being specified by ${\para}$ and ${\perp}$, respectively, 
and  fermion and boson thermal frequencies, $\varepsilon_{n}=(2n+1)\pi /\beta$ and  
$\omega_{n}=2n\pi /\beta$, respectively.

The lowest-order   scaling equation for $V^{\rm SCd}$ is depicted in 
Fig.~3(d),  and is written as
\begin{eqnarray}
{dV^{\rm SCd}(l)\over dl}
&=&-\left[\tilde t_{\perp}(l)g^{\rm SCd}(l) \right]^{2}
\non\\
&+&
2g^{\rm SCd}(l) V^{\rm SCd} (l)
-{1\over 2} \left[V^{\rm SCd}(l)\right]^{2},  \label{eqn:scalingV}
\end{eqnarray}
 where  the transverse momentum transfer of the pair is set at $q_{\perp}=0$.
The coupling for the SCd  pair field is given by
$g^{{\rm SCd}}={1\over 2}(g^{(1)}_{t}+g^{(2)}_{t}-g^{(1)}_{0}-g^{(2)}_{0})$.
 By lengthy but straightforward manipulation, we obtain   similar scaling 
equations for all  $V^{M}_{\mu}(l)$.
We have solved  them with the initial conditions
\begin{equation}
V^{M}_{\mu}(0)=0
\end{equation}
 and confirmed that  $V^{\rm SCd}$ always dominates the other channels. 
This situation is quite reasonable on the physical
grounds that the interladder pair tunneling stabilizes the most dominant intraladder correlation, i.e, $d$-wave superconducting correlation.
Below, we  focus on  the $d$-wave superconducting channel.
The third term of the r.h.s of eq.(\ref{eqn:scalingV}) causes divergence of  ${ V}^{{\rm 
SCd}}$ at a 
critical scaling parameter $l_{c}$ determined  by
\beq
{ V}^{{\rm SCd}}(l_{c})=-\infty\label{eqn:defoflc2}
\eeq
In the lower half planes of Figs.~4(b) and 4(c), we show the scaling flows 
of $V^{{\rm SCd}}(l)$ for  $\tilde t_{\perp0}=0.01$ and $0.04$.

Figures 4(b) and 4(c) show that, in the 
case of $\tilde t_{\perp0}=0.01$, only the two-particle crossover occurs, while
 in the case of $\tilde t_{\perp0}=0.04$, the 
one-particle crossover dominates the two-particle crossover.
We  identify the scaling parameter with the absolute temperature   as
$
l=\ln {E_{0}\over T}.
$
Thus, based on eqs.(\ref{eqn:defoflc1}) and (\ref{eqn:defoflc2}), we   define 
the one-particle {\it crossover temperature}, $T_{\rm cross}$, and the $d$-wave superconducting 
{\it transition temperature}, $T_{c}$, as
\beq
\begin{array}{c}
T_{\rm cross}=E_{0}e^{-l_{\rm cross}},\\
T_{c}=E_{0}e^{-l_{c}}.
\end{array}
\eeq 
$T_{\rm cross}$ and $T_{c}$ correspond to $T_{x1}$ and $T_{x2}$, respectively,  introduced by Bourbonnais and Caron\cite{Bourbonnais}
for weakly coupled chains.

By solving the scaling equations for various $\tilde  t_{\perp0}$ with the 
fixed value of the Hubbard repulsion,  $\tilde U=0.3$,
we obtained the phase diagram of the system in terms of $\tilde t_{\perp0}$ and the reduced
 temperature $\tilde T=T/E_{0}$, as shown in Fig.~5.
\begin{figure}
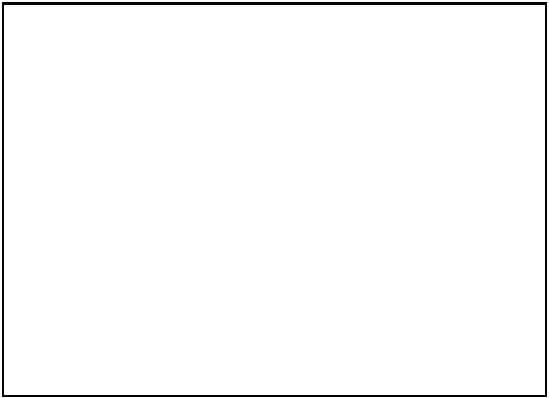

\begin{center}
\figureheight{5cm}
\caption{Phase diagram of the weakly coupled Hubbard ladder system.
{\bf SGM}, {\bf SCd} and {\bf 2D} denote the sping gap metal phase,
the $d$-wave superconducting phase and the two-dimensional phase, respectively.
 The interladder one-particle hopping and
temperature are  scaled by the initial bandwidth cutoff, $E_{0}$.}
\end{center}
\end{figure}
Roughly speaking, we can regard increasing $\tilde t_{\perp0}$ as applying the pressure
 under which the bulk superconductivity was
actually observed.
We found that  there exists a crossover value of the interladder 
one-particle hopping, $\tilde t_{\perp c}\sim 0.025$.

For $0<\tilde t_{\perp0}<\tilde t_{\perp c}$,  the phase transition into the $d$-wave superconducting (SCd) phase  occurs
 at a finite transition temperature, $T_{c}$,  via the two-particle crossover.
In the temperature region, $T<T_{c}$, coherent   Josephson tunneling of the Cooper pairs in the interladder transverse direction occurs.
Here we have to be careful about identification of the finite temperature phase above $T_{c}$, where the system is in {\it the isolated ladder regime.}
As the temperature scale decreases, the isolated ladders are {\it gradually} scaled toward their low-energy asymptotics, 
the  SGM phase.
The gradual change of the density of darkness in the SGM phase in Fig.5 schematically depicts this situation.
The SGM phase is characterized by the strong coupling values of intraladder couplings, $g_{t}^{(1)}=g_{t}^{(2)}=1$ and $g_{0}^{(1)}=-1$, (see (\ref{FP})).
The critical scaling parameter, $l_{c}$, is in the region,
$l_{c}>5.3$,  around which the intraladder coupling constants almost reach thier fixed point values (see Fig.4.(a)).
Thus  we expect that  the spin gap is well developed near  $T_{c}$.
Within the framework of the PRG approach,
we cannot say for certain whether the spin gap survives in the SCd phase or not.

For $\tilde t_{\perp c}<\tilde t_{\perp0}$, around the crossover temperature $T_{\rm cross}$, the system crosses over to the  2D phase
via the one-particle crossover.
The crossover value of the scaling parameter, $l_{\rm cross}$, is in the region
$l_{\rm cross}<4.6$, around which the intraladder coupling constants are far away from  thier fixed point values (see Fig.4(a)).
Thus it is disputable to  assign the phase above $T_{\rm cross}$ to the SGM phase.
In the temperature region, $T<T_{\rm cross}$, the interladder {\it coherent} band motion occurs. Then the physical properties of the 
system would strongly depend on the shape of the 2D Fermi surface.

Finally we briefly compare  the present case with the case of  coupled  
chains within the  PRG scheme.\cite{Bourbonnais,Boies}  The scaling 
equation for the interchain one-particle hopping, 
instead of (\ref{eqn:RGfort}), gives 
$T_{\rm cross}\sim E_{0}\left[t_{\perp0}/E_{0}\right]^{1/(1-\theta)}$,\cite{Bourbonnais,Boies}
where one has the exact result for the anomalous exponent, $\theta\leq 1/8$, 
for the non-half-filled Hubbard model.\cite{Schulz2}
Consequently  $t_{\perp}$ becomes {\it always relevant}, contrary to the coupled ladder case.
This difference reflects the fact that the isolated Hubbard ladder belongs to the 
strong coupling universality class (SGM phase), while the isolated Hubbard chain 
belongs to the weak coupling (Tomonaga-Luttinger) universality class.
In that sense,  we can conclude that {\it the spin gap opening
strongly suppresses the one-particle crossover so that 
the $d$-wave superconducting transition via the two-particle crossover 
is strongly assisted in the weakly coupled ladder system.} 
\section*{Acknowledgements}
J.K was   supported by a Grant-in-Aid for Encouragement for Young Scientists   from the Ministry of Education, Science, Sports and Culture, Japan. 


\begin{thebibliography}{999}
\bibitem{comment} We ignore the misfit of the rungs in the neighboring ladders  which  exists  in real Sr$_{14-x}$Ca$_{x}$Cu$_{24}$O$_{41}$
compounds, since   we  take the continuum limit along the leg and 
consequently the misfit  never appears in the present theory.
\end{thebibliography}
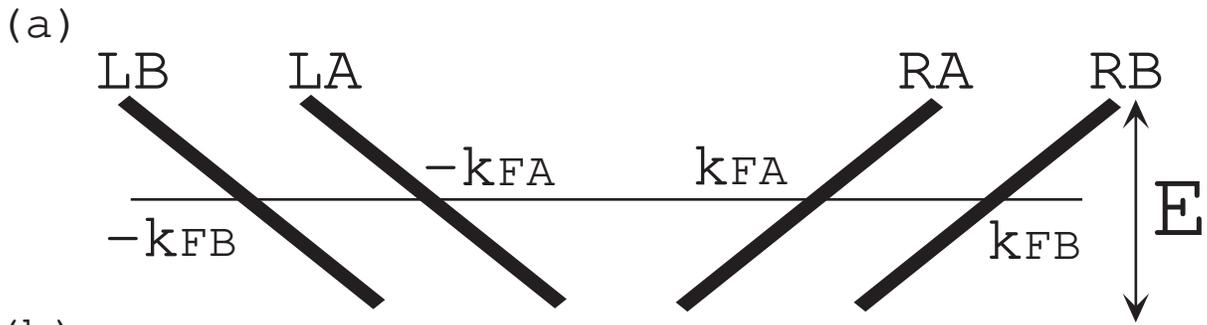
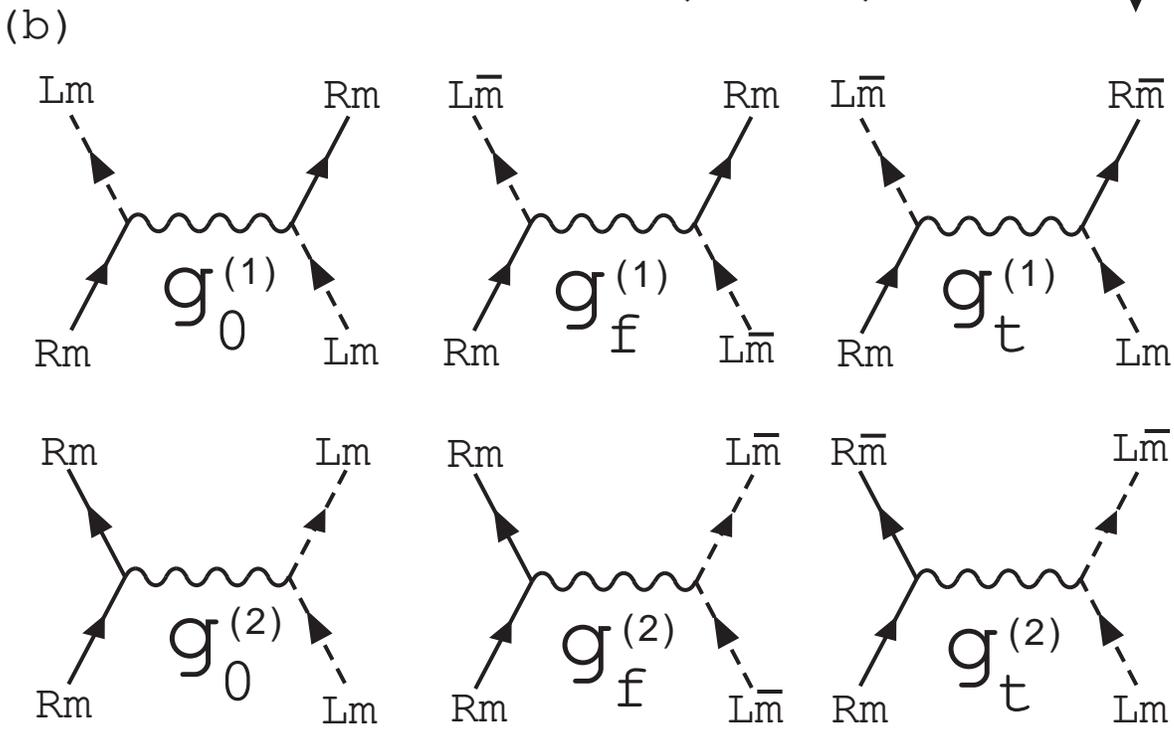
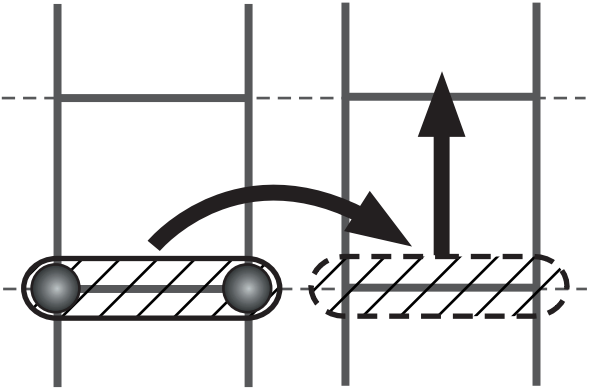
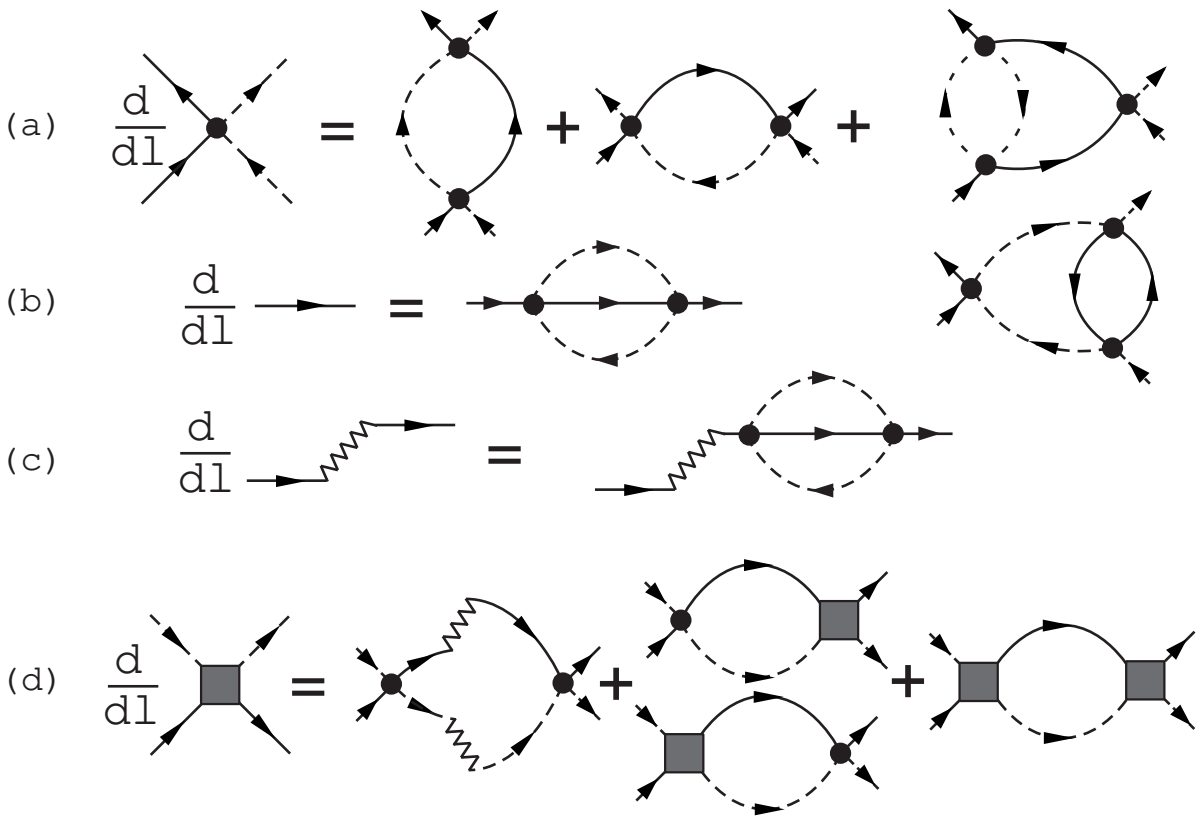
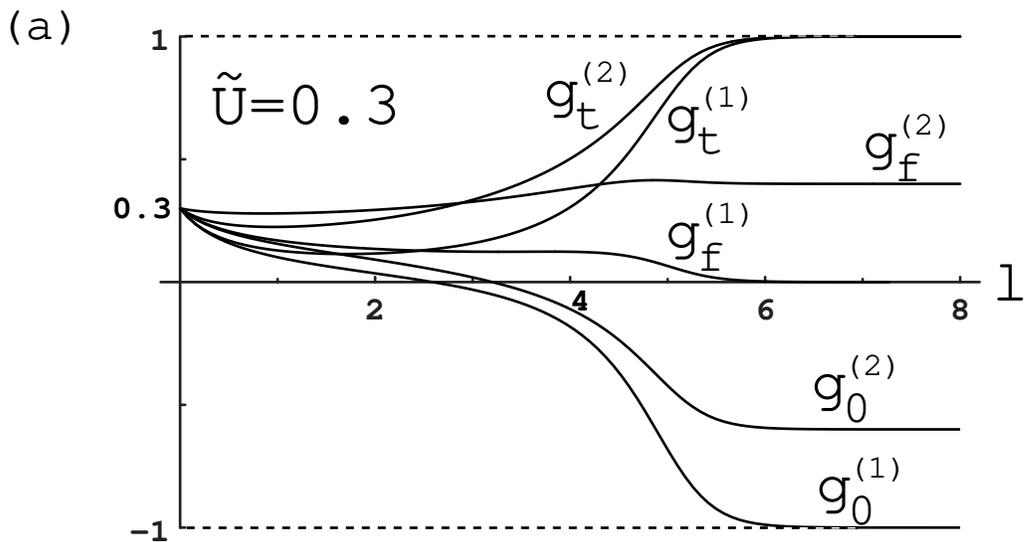
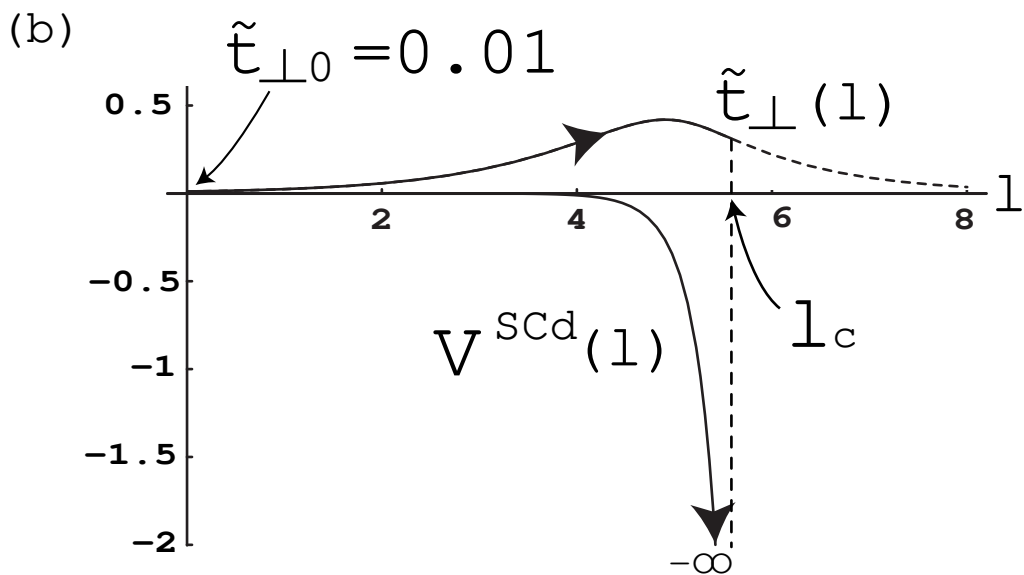
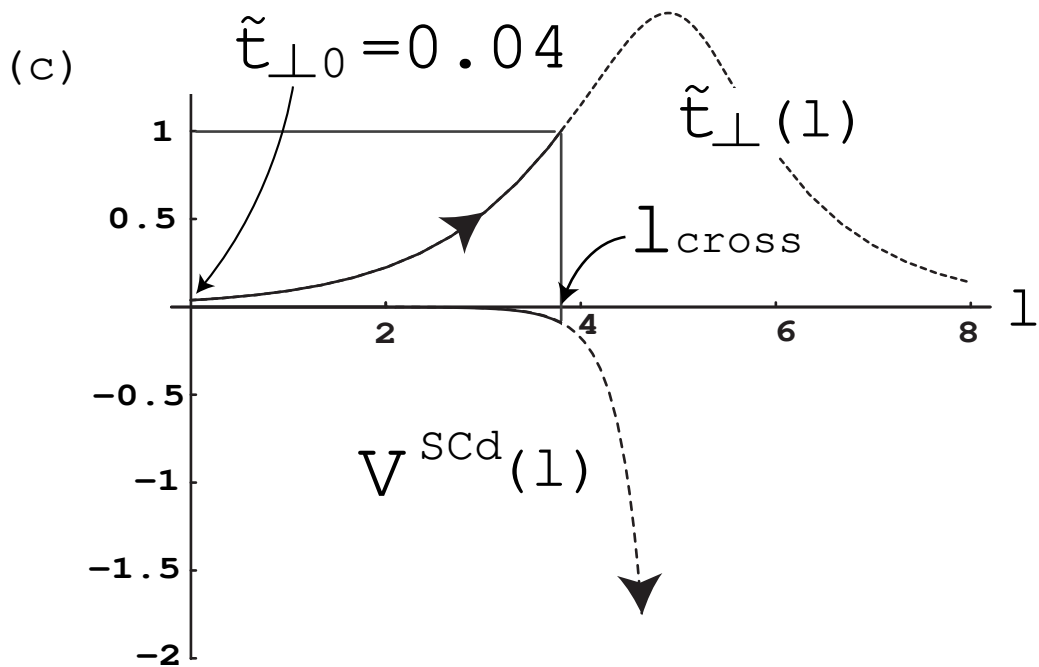
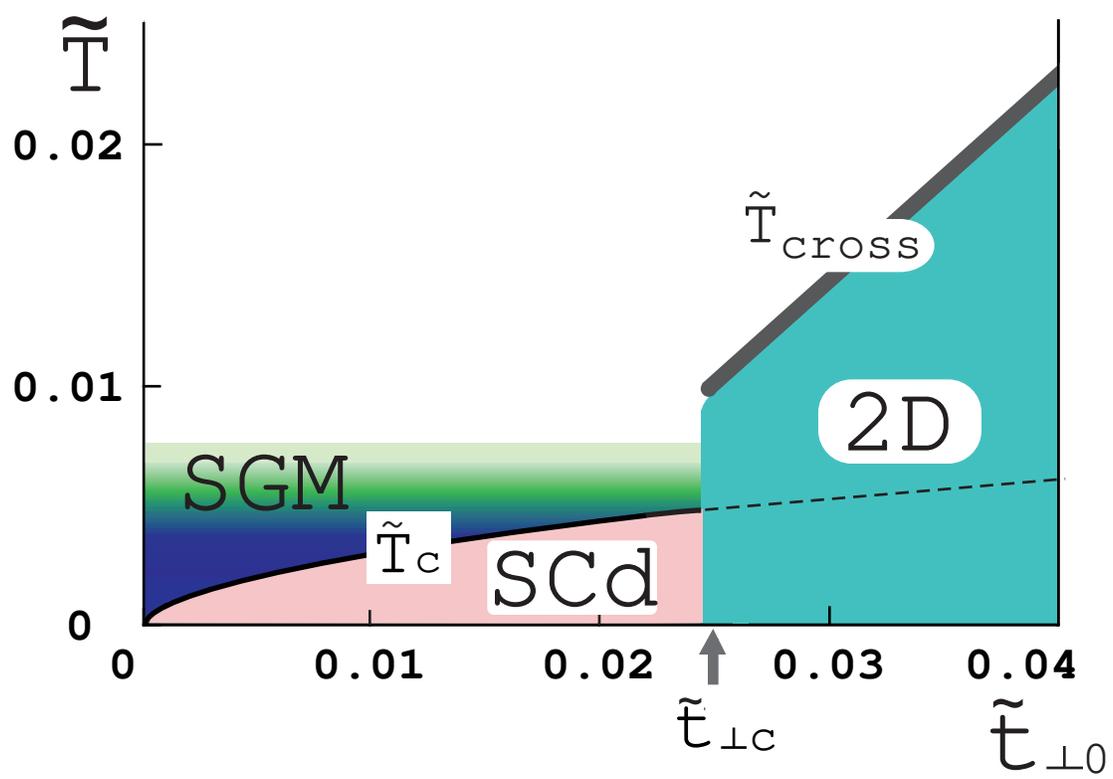
\end{document}